\documentclass[11pt]{article}
\newcommand{\version}{January 5, 2004}

\usepackage{amsthm,amsfonts,amsmath}
\swapnumbers
\theoremstyle{plain}

\theoremstyle{definition}

\theoremstyle{remark}
\newcommand{\upchi}{\raise1pt\hbox{$\chi$}}
\newcommand{\R}{{\mathord{\mathbb R}}}

\newcommand{\C}{{\mathord{\mathbb C}}}

\numberwithin{equation}{section}
\begin{document}

\markboth{\scriptsize{LL \version}}{\scriptsize{LL \version}}

\title{\bf{A Note on Polarization Vectors in Quantum Electrodynamics}}
\author{\vspace{5pt} Elliott H. Lieb$^1$ and
Michael Loss$^{2}$ \\
\vspace{-4pt}\small{$1.$ Departments of Physics and Mathematics, Jadwin
Hall,} \\
\small{Princeton University, P.~O.~Box 708, Princeton, NJ
  08544}\\
\vspace{-4pt}\small{$2.$ School of Mathematics, Georgia Tech,
Atlanta, GA 30332}  \\ }
\date{\version}
\maketitle

\footnotetext
[1]{Work partially
supported by U.S. National Science Foundation
grant PHY 01-39984.}
\footnotetext
[2]{Work partially
supported by U.S. National Science Foundation
grant DMS 03-00349.\\
\copyright\, 2003 by the authors. This paper may be reproduced, in its
entirety, for non-commercial purposes.}

\begin{abstract}
  A photon of momentum $k$ can have only two polarization states, not
  three.  Equivalently, one can say that the magnetic vector potential
  $A$ must be divergence free in the Coulomb gauge. These facts are
  normally taken into account in QED by introducing two polarization
  vectors $\varepsilon_\lambda(k)$ with $\lambda \in \{1,2\}$, which
  are orthogonal to the wave-vector $k$. These vectors must be very
  discontinuous functions of $k$ and, consequently, their Fourier
  transforms have bad decay properties. Since these vectors have no
  physical significance there must be a way to eliminate them and
  their bad decay properties from the theory. We propose such a way
  here.
\end{abstract}

\section{Introduction} \label{intro}

In quantum electrodynamics it is  necessary to choose a gauge for
the electromagnetic field, and we shall use the Coulomb gauge here for
the reason that it is the only gauge in which one gets the correct
electromagnetic field and electromagnetic energy by a minimization
principle. This comes from the fact that the magnetostatic
interaction energy of a current distribution can be found by
minimizing $(8\pi)^{-1} \int B^2 - \int j \cdot \mathbf{A}$ with respect
to the vector potential whereas the (positive) electrostatic
interaction energy of a charge distribution $\rho$ is not
given by the minimum of ${8\pi}^{-1} \int |\nabla \phi|^2 -\int \phi \rho$.
(For a discussion of this see, e.g., \cite{L}.)
In the Coulomb gauge the electrostatic part of the interaction among
particles is given directly by Coulomb's law $e_ie_j/|x_i-x_j|$.  The
curl-free part of the electric field is not a dynamical variable in
this gauge.

The dynamical field, whose dynamics is `quantized', is the magnetic
field. The (ultraviolet cutoff) magnetic vector potential
is customarily defined by
\begin{equation}\label{apot}
\mathbf{A}(x) = \frac{\sqrt{\hbar c}}{2\pi} \sum_{\lambda=1}^2 \int_{\R^3}
\frac{\varepsilon_\lambda(k)}{\sqrt{|k|}} 
\widehat\chi_{\Lambda}^{\phantom{B}}(k) \left(
a^{{\phantom{\ast}}}_\lambda(k) e^{ik\cdot x} +
a_\lambda^{\ast}(k) e^{-ik\cdot x}\right)dk
\  ,
\end{equation}
where the function $\widehat\chi_{\Lambda}^{\phantom{B}}$ is a
radial function in $k$ space that vanishes outside the ball whose
radius is the ultraviolet cutoff $\Lambda$. 
The creation and annihilation
operators of photons of momentum $k$ and polarization $\lambda$, 
$a_\lambda (k)$ and $a_\lambda^* (k)$, which act on
Fock-space (over $L^2(\R^3) \otimes  \C^2$),
satisfy the canonical commutation relations
\begin{equation}
[a_\lambda(k), a^{\ast}_{\nu} (q)] = \delta ( k-q)
\delta_{\lambda, \nu}\ , ~~~ [a^{{\phantom{\ast}}}_
{\lambda}(k), a^{{\phantom{\ast}}}
_{\nu} (q)] = 0, \quad {\rm{etc}}\ .
\end{equation}

The magnetic field is $\mathbf{B}(x) = {\mathrm{curl}\ }\mathbf{A}(x)$.
The vectors $\varepsilon_\lambda(k)$ are two orthonormal polarization
vectors, which are perpendicular  to $k$ as well as to each other.

The field energy, $H_f$, sometimes called $d\Gamma(\omega)$, is given
by
\begin{equation}\label{eq:fielden}
H_f = \hbar c \sum_{\lambda=1,2} ~ \int_{\R^3} ~ |k| ~
 a^*_\lambda(k)  a_\lambda (k) d k \ . 
\end{equation} 
There is no cutoff in $H_f$. The energy of a photon is $\hbar c |k|$.

The polarization vectors are necessarily discontinuous. They must be
discontinuous on every sphere centered at the origin, in fact, for the
well known reason that ``one cannot comb the hair on a sphere''.  One
possible choice, but by no means the only one, is
\begin{eqnarray}\label{polvec}
\varepsilon_1(k) &=& \frac{(k_2, -k_1, 0)}{\sqrt{k_1^2+k_2^2}} \ ,
\nonumber \\ 
\varepsilon_2(k) &=& \frac{k}{|k|} \wedge  \varepsilon_1(k) \ .
\end{eqnarray} 

While polarization is physically meaningful and measurable, the
polarization vectors and the corresponding operators $a^\#_\lambda(k)$
have no direct physical meaning. The polarization \textit{vectors}
merely form an arbitrarily chosen and hence \textit{unobservable} basis 
for vectors perpendicular to 
$k$. It should be possible to
define the theory without the unphysical operators $a^\#_\lambda(k)$.

\textit{Why does this matter?} For an atom in the ground state $\Phi$ one
expects that the photon density decays towards zero with increasing 
distance from the nucleus.
Since all the relevant quantities of the radiation field are expressed in terms
of operators that act on $k$ space, it is convenient to establish this decay
by finding a bound on
\begin{equation}
\sum_{\lambda=1}^2 \int \Vert \nabla_k a_\lambda(k) \Phi \Vert^2 dk \ ,
\end{equation}
noting that smoothness in $k$ space translates into decay in 
configuration space. To calculate the above expression one needs
to compute 
$\nabla_k\  \varepsilon_\lambda (k)$ whose singularity causes needless 
complications, as in \cite{G,GLL}.

A related complication caused by the polarization vectors occurs if one wants to
calculate the coupling function $h^i_\lambda(x)$ that measures the strength
of the interaction of the electron with the photon field. This function is
given by the Fourier transform of the $k$-dependent quantity appearing
in (\ref{apot}), namely (with superscripts $i=1,2,3$ denoting components)
\begin{equation}\label{couplingfunction}
h^i_\lambda (y)=\frac{1}{2\pi} \int
\frac{\widehat\chi_{\Lambda}^{\phantom{B}}(k)} {\sqrt{|k|}} 
\varepsilon^i_\lambda (k) e^{-ik
\cdot y} dk
\end{equation}

The reason for the interest in $h^i_\lambda(y)$ is that in some problems,
such as the verification of the binding condition for atoms \cite{atom}
or the existence of the thermodynamic limit,
it is necessary to localize the electromagnetic field in $x-$space.
Thus, if we formally define 
\begin{equation}
\check a_\lambda(x) = \frac{1}{(2\pi^3)}\int_{\R^3}a_\lambda(k) e^{ik\cdot x}
dk 
\end{equation}
to be the Fourier transform of
the operator $a_\lambda(k)$ then
\begin{equation}\label{xspacea}
A^i(x) = \sum_{\lambda =1}^2 \check a_\lambda (h^i_\lambda( x- \cdot)
+ \check a_\lambda^* (h^i_\lambda( x- \cdot) \ .
\end{equation}
(Formally, $\check a_\lambda (h^i_\lambda( x- \cdot)= \int \check a_\lambda (y)
h^i_\lambda( x- y) dy$.) 
We want $h^i_\lambda$ to have a rapid fall-off in order that 
$\check a_\lambda(y)$ be localized with $y$ as close to $x$ as possible.

Unfortunately, because of the discontinuity of $\varepsilon^i_\lambda (k)$ it is very
difficult to decide what  the fall-off of $h^i_\lambda(y)$ is. The decay, being slow
in directions that are perpendicular to the direction of the singularity of
$\varepsilon^i_\lambda (k)$, will be nonuniform.
With a smooth cutoff function $\widehat{\chi}_\Lambda (k) $ we can, by a suitable choice of the
polarization vectors, make $h^i_\lambda(y)$ decay  in such a way
that $\int |x|^{2\gamma} |h^i_\lambda(y)|^2 dy $ is finite for all
$\gamma <1$.  This weak form of fall-off is useful, but 
inconvenient to work with. Morover, one can envision situations where the
nonuniformity of the decay will cause problems.

In contrast, if we omit the
$\varepsilon^i_\lambda (k)$ then the fall-off of the basic coupling function 
\begin{equation}\label{couplingfunctiongood}
h (y)=\frac{1}{2\pi} \int
\frac{\widehat\chi_{\Lambda}^{\phantom{B}}(k)} {\sqrt{|k|}} 
 e^{-ik
\cdot y} dk \ ,
\end{equation}
will be $|y|^{-5/2}$ as $|y|\to \infty$ if
we choose the cutoff $\widehat \chi_{\Lambda}^{\phantom{B}}(k)$ to be
smooth.  To see this we note that $|k|^{-1/2}$ is the Fourier transform of $|y|^{-5/2}$
in the sense of distributions \cite[Theorem 5.9]{Analysis}. The
Fourier transform of $\widehat\chi_{\Lambda}$ is real analytic and
decays faster than any inverse power of $|y|$.  Hence, the convolution
of $\chi_\Lambda$ (the Fourier transform of $\widehat\chi_{\Lambda}$),
with $|y|^{-5/2}$ decays like $|y|^{-5/2}$.  As an aside we note that a sharp ultraviolet
cutoff, $h(y)$ would decay only like $|y|^{-2}$, which turns out to be
insufficient for a good localization of the photon states.  

In an attempt to eliminate the polarization vectors from the formalism
it was suggested in \cite{atom} that it would be better to start with
a formalism that contains only ``divergence-free" vector fields as the
dynamical variables.  In particular, the Fock space would be built
over the $L^2$-space of divergence-free vector fields instead of
$L^2(\R^3)\otimes \mathbb{C}^2$. While this approach removes the arbitrariness
in the choice of polarizations it causes problems when one attempts to
localize photons. If one multiplies a divergence-free vector field $v$
by a smooth cutoff function $j$ the result is not a divergence-free vector
field in general. One possible localization procedure is to write $v={\mathrm{curl}\ } w$ and then use the
field $j v + \nabla j \wedge w$. Again, this is fairly tedious.

Since then, we have found an easier  way and
that is the subject of this paper.

\section{The Third Photon Mode}

Our proposal is really very simple. Let us introduce
three photon degrees of freedom for each $k\in \R^3$
namely, 
\begin{equation}
\mathbf{a}(k) =\{a^1(k), \ a^2(k), \ a^3(k)\}, 
\end{equation}
with the $a^i$ satisfying canonical commutation relations
$\left[a^i(k),\ a^j(q)\right] =0$ and 
$\left[a^i(k),\ a^{j*}(q)\right] 
= \delta_{i,j} \delta(k-q)$.
Another way to say this is that we use the Fock space
over $L^2(\R^3)\otimes \mathbb{C}^3$.
The field energy involves all three modes:
\begin{equation}\label{fieldennew}
H_f = \hbar c \sum_{j=1}^3 ~ \int_{\R^3} ~ |k| ~
a^{j*}(k) a^j(k) d k \ . 
\end{equation}

The vector potential is replaced by
\begin{equation}\label{apotnew}
\mathbf{A}(x) = \frac{1}{2\pi} \int_{\R^3}
\frac{1}{\sqrt{|k|}} 
\widehat\chi_{\Lambda}^{\phantom{B}}(k) \frac{k}{|k|}\wedge \left(
\mathbf{a}(k) e^{ik\cdot x} +
\mathbf{a}^{\, \ast}(k) e^{-ik\cdot x}\right)dk
\  ,
\end{equation}
and the analogue of (\ref{xspacea}) is the vector equation
\begin{equation}\label{Anew}
\mathbf{A}(x) =  i\ \textrm{curl\ } \check{\mathbf{a}} (\widetilde{h}( x- \cdot))
-i\ \textrm{curl\ } \check {\mathbf{a}}^{\, *} (\widetilde{h}( x- \cdot)) \ .
\end{equation}
The function $\widetilde{h}$ is related to (\ref{couplingfunctiongood})
\begin{equation}\label{couplingfunctionnew}
\widetilde{h}(y)=\frac{1}{2\pi} \int
\frac{\widehat\chi_{\Lambda}^{\phantom{B}}(k)} {|k|^{3/2}} 
 e^{-ik \cdot y} dk \ .
\end{equation}

Note that 
\begin{equation}\label{note}
\left \{ \textrm{curl}\ \check{\mathbf{a}} (\widetilde{h}( x- \cdot))^{\phantom{B}}
\right \}^\ell
= \sum_{j=1}^3 \left\{ \check a^j(\partial_i \widetilde{h}(x-\cdot))
- \check a^i(\partial_j \widetilde{h}(x-\cdot))\right\}   \varepsilon_{ij\ell}\ .
\end{equation}

Following the argument about the fall-off of $h(y)$, we see that the
fall-off of $\widetilde{h}(y)$  is bounded by
$|y|^{-3/2}$ and its derivatives are bounded by
$|y|^{-5/2}$. Thus, the desired fall-off   $|y|^{-5/2}$ is
obtained, using (\ref{note}),  for the localization of $\mathbf{A}(x)$.

We have the following situation: The quantized field operators have
been increased from two to three for each $k-$value. Nevertheless, the
vector potential $\mathbf{A}$ (which mediates the interaction of matter and
radiation) still has the property that its $k-$Fourier transform is
perpendicular to $k$. The field energy has been supplemented by an
additional mode (which we might think of as 'dark energy' since it
does not interact with matter and is, therefore, not detectable). For reasons
outlined at the beginning of this paper these additional modes 
are not longitudinal photons. Those have been eliminated from the theory
by choosing the Coulomb gauge. 

What
we shall show next is that this theory, in which the Hamiltonian
describing matter and its interaction with radiation (relativistic or
non-relativistic) is formally the same except for the extra invisible
mode, gives the same physics as the old theory with the two
polarization vectors.

Before continuing, let us note that there is an alternative to
(\ref{apotnew}) that accomplishes the same thing. We can replace the
vector $|k|^{-1}k\wedge \mathbf{a}(k)$ in (\ref{apotnew}), whose
$\ell$-component is $|k|^{-1}k^i \ a^j(k) \ \varepsilon_{ij\ell}$, by
the vector whose $\ell$-component is
\begin{equation} \label{proj}
\sum_{\lambda =1}^2 \sum_{j=1}^3 \varepsilon_\lambda ^\ell(k)\,  
\varepsilon_\lambda ^j(k) \, a^j(k) = 
\left\{ \sum_{\lambda =1}^2 \varepsilon_\lambda (k) \, \varepsilon_\lambda (k)\cdot
\mathbf{a}(k) \right\}^\ell
= a^\ell(k) - \frac{k^\ell}{|k|^2}\,  k\cdot \mathbf{a}(k) \ .
\end{equation}
The vector whose $\ell$-component is displayed in (\ref{proj}) is the
projection of $\mathbf{a}$ onto the plane perpendicular to the vector $k$.
The vector $|k|^{-1}k\wedge \mathbf{a}(k)$ is a vector that is perpendicular to
both $\mathbf{a}$ and $k$ and whose norm is the same as the vector in 
(\ref{proj}). 

The theories with (\ref{proj}) and with $|k|^{-1}k\wedge \mathbf{a}(k)$ are
indistinguishable. We shall continue with (\ref{apotnew}).

\section{Equivalence of the Two Theories}

Let us start with the theory defined by the vector potential
(\ref{apotnew}) and field energy $H_f$ in (\ref{fieldennew}). That is
to say, the total Hamiltonian of matter plus radiation has the form
\begin{equation}\label{totalham}
H_{\mathrm{total}} = H_{\mathrm{matter}}(\mathbf{A}) + H_f
\end{equation}
where $H_{\mathrm{matter}}(\mathbf{A})$ describes the matter (as particles, or
as a quantized field, relativistic or non-relativistic). The important
point is that it depends on the radiation field only through the
quantized vector field $\mathbf{A}$ in (\ref{apotnew}).

Instead of the three $k$-dependent operators $\mathbf{a}$ we introduce
\begin{equation}
a_\lambda(k) = \varepsilon_\lambda(k)\cdot \mathbf{a}(k), \qquad\qquad
a_0(k) = \frac{k}{|k|}\cdot \mathbf{a}(k)\ 
\end{equation}
and check that (with $\nu= \lambda$ or $\nu=0$)
\begin{equation}\label{commnew}
\left[a_\nu (k), a_\mu(q)\right] =0 \qquad 
\left[a_\nu(k), a_\mu^*(q)\right] = \delta_{\nu, \mu} \delta(k-q) \ .
\end{equation}

We also observe that $\mathbf{A}$ can be written using only the two
$a_\lambda(k)$ as in (\ref{apot}) and that the field energy
(\ref{fieldennew}) is
\begin{equation}
H_f=  \hbar c \, \int_{\R^3} |k| \left\{\sum_{\lambda =1}^2 a^*_\lambda(k)
a_\lambda(k) + a^*_0(k) a_0(k)\right\}\, dk \ .
\end{equation}

All of this is breathtakingly elementary to verify. The conclusion,
however, is interesting. This `three-component' model
(\ref{totalham}), which is simpler to deal with than the usual
`two-component' model because the polarization vectors are absent, has
the property of merely describing the `two-component' theory plus one
totally independent scalar field $a_0$ whose time evolution is
governed by the Hamiltonian
\begin{equation}
H_0 = \hbar c \, \int_{\R^3} |k| a^*_0(k) a_0(k)\, dk \ .
\end{equation}
Thus, the (Heisenberg) time evolution of $a_0$ is simply
$a_0(k,t) = a_0(k) e^{i c |k| t}$. 

The eigenvalues of (\ref{totalham}), on the other hand, are those of
the original  `two-component' model plus the scalar field energy, whose
eigenvalues (if the radiation field is enclosed in a finite box) are of the form
\begin{equation}
\hbar c \sum |k|\, n_k \ .
\end{equation}
The $n_k$ are nonnegative integers, of which only a finite
number are positive. For the ground state we choose all $n_k= 0$.
The energy in the scalar field mode is not observable.

If we are interested in the thermodynamic limit of ordinary matter
coupled to the radiation field in a positive temperature Gibbs state
we have to proceed as follows. First, we imagine the universe to be a
huge box of volume $\mathcal{V}$, while the matter is confined to a
much smaller box of volume $V$. We would have to imagine this, even if
we stayed with the usual `two-component' formulation. Then we would
take the limit $\mathcal{V}\to \infty$, after subtracting the well
known positive temperature free energy $F(T)$ of the field, whose asymptotic 
value was  calculated by Planck in 1900 \cite{planck}.  
\begin{equation}\label{free}
\frac{F(T)}{k_{\mathrm{Boltzmann}}T} =  - 3  \frac{\mathcal{V}}{(2\pi)^3}
\int_{\R^3} \log
\left( 1-\exp\left[- \frac{\hbar c |k|} 
{k_{\mathrm{Boltzmann}} T}\right] \right)  \, dk
\end{equation}

\medskip \noindent
This subtraction is necessary in either
theory, the only difference being the subtraction of the scalar field
energy in our `three-component' theory. Hence the factor 3 instead of 2 in
(\ref{free}). After this $\mathcal{V}\to
\infty$ limit, one takes the usual ${V}\to \infty$ limit. In the end,
the scalar field contributes nothing. Its role is only to contribute
some simplification to a difficult calculation.

\end{document}